# On the mass determination in liquid utilizing measurement of only the fundamental flexural resonances of the micro-/nanomechanical based mass sensors

Ivo Stachiv

*Department of Mechanical Engineering, National Kaohsiung University of Applied Sciences, Kaohsiung, Taiwan*

*Institute of Physics, Czech Academy of Sciences, Prague, Czech Rep.*

(E-mail address: stachiv@fzu.cz)

**Abstract**

Micro-/nanomechanical mass sensors are capable to quantitatively determine molecule mass from only first three (two) measured cantilever (bridge) resonant frequencies. However, in liquid solutions that are relevant to most of the biological systems, the mass determination is challenging because the $Q$-factor due to fluid damping decreases and, as a result, usually just the fundamental resonant frequencies can be correctly identified. Moreover, for higher modes the resonance coupling, noise and internal damping have been proven to strongly affect the measured resonant frequencies and, correspondingly, the accuracy of the estimated masses. Here, we derive the easy accessible expressions enabling the quantitative mass(es) determination just from the fundamental resonant frequencies of the micro/nanomechanical mass sensor under intentionally applied axial tension, which can be easily created and controlled by the electrostatic or magnetostatic forces. We also show that typically achievable force resolution has a negligible impact on the mass determination and the mass sensitivity.

## I. INTRODUCTION

Vibrating micro-/nanomechanical beams are one of the main component used in nanotechnology for detection of various physical quantities including pressure, force,[1] quantum state,[2] spin,[3] thin film mechanical properties[4] and molecule masses.[5] Particularly, the micro-/nanomechanical mass sensors possess the ultrahigh sensitivity, excellent selectivity, the operating frequencies up to several gigahertz with the extraordinary controllability via optomechanical or electromechanical coupling and, finally, enable real time mass spectrometry with capability to reach the ultimate limits of mass detection.[6,7] These devices usually measure shift of the flexural resonant frequencies caused by the attached molecule(s). Resonant frequencies decrease when the molecule(s) or nanoparticle(s) is (are) attached on the resonator surface and value of the frequency shift depends on the attached mass(es) and the position(s) of attachement.[8,9] It has been shown that by measuring multiple vibrational modes: two for bridge and three for cantilever, single particle mass can be unambiguously determined.[10] Recently, this approach allowed to achieve even a single-protein real time mass spectrometry in vacuum.[7]

In general, flexural motion can be described by a well-known Euler-Bernoulli beam equation.[11] This equation, however, predicts accurately usually just the fundamental resonant frequencies, while for higher modes the accuracy of predicted resonances reduces due to either the internal friction losses[11] or the existence of coupling between out- and in-plane flexural, torsional and longitudinal oscillations.[12] Furthermore, in aqueous solutions that are relevant to most of the biological systems, dissipation due to surrounding media dominates causing hence a large decrease of the quality factor ($Q$-factor) and, consequently, usually just the fundamental resonant frequencies can be correctly identified.[13] Exceptionally even the first two modes can be detected.[14] As a result, in gases and liquids including the physiological solutions, the attached molecule(s) are evaluated either from the measured shift of the flexural fundamental mode by accounting for the uniform distribution of the added molecules over the entire resonator length[15] or from measured shifts of the flexural and longitudinal (torsional) fundamental modes.[16] Former method does not account for the molecule(s) attachment position(s), therefore the one does not allow the quantitative single / multiple mass determination. The latter method is capable of single mass spectrometry in fluid but, nevertheless, it is still highly challenging to simultaneously determine fundamental flexural and longitudinal / torsional[17] vibrational modes.

Thus it is evident from the above discussion that the method capable of the quantitative mass determination from just measured fundamental flexural resonances, which is of practical importance in many scientific fields including chemistry, biology and medicine, is still missing. In response, here we extend previous works[8-10,16,18-23] carried out on the vibrating cantilever and suspended micro-/nanomechanical based mass sensors (see Fig. 1) to show that the quantitative determination of the attached molecule masses can be realized just from measured shifts of the fundamental flexural frequencies under different intentionally applied axial tensile force $F_T$. We must emphasis here that for majority of the nanomechanical based mass sensors the flexural vibrations are realized and controlled via an external electrical (or electromagnetic) field,[5,18,19] which indeed itself creates tensile force. Even the mechanical generation of $F_T$ has been reported.[23]

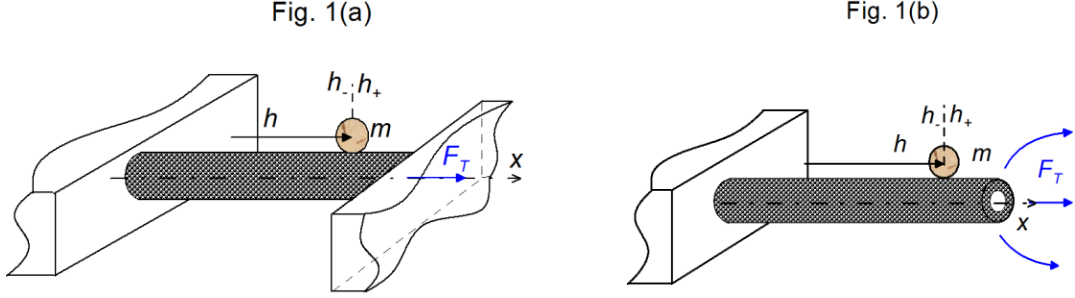

Fig. 1. A schematic representation of the nanomechanical based mass sensor under axial tensile force with an attached particle in (a) bridge and (b) cantilever configurations.

## II. BACKGROUND THEORY AND METHOD OF QUANTITATIVE MASS(ES) DETERMINATION FROM SHIFTS OF ONLY THE FUNDAMENAL MODE

To begin we recall that the aggregate mass of the attached molecules or particles $m_\Sigma$, where $m_\Sigma = \sum_{i=1}^{N} m_i$ and $N$ stands for the number of attached masses, is considered to be essentially smaller than the resonator mass $M$, i.e. $m_\Sigma \ll M$.[8-10,16] Then following the work of Stachiv et al.[10,16] and accounting for $N$ matching conditions the frequency shift of axially loaded nanomechanical based mass sensor caused by the $N$ attached masses can be obtained in the following way

$$(f_0 - f)/f_0 = 2\varepsilon_\Sigma \alpha_\Sigma(h_j^*, \gamma_0), \qquad (1)$$

where $f_0$ and $f$ are the unloaded and loaded by $N$ masses resonant frequencies; $\varepsilon_\Sigma = m_\Sigma/M$, $\alpha_\Sigma(h_j^*, \gamma_0)$ is the position function given in Appendix, $h_j^* = h_j/L$ are the dimensionless attachment positions, i.e. $j \in \{1, \ldots, m_N\}$; $L$ is the resonator length and $\gamma_0^2$ are dimensionless resonant frequencies of the unloaded resonator obtained as a solution of the appropriate transcendental equation.[21,22]

Similarly, for energy approach the frequency shift caused by $N$ attached masses can be easily obtained in the same manner as given in Refs. 8 and 9 and it reads

$$f/f_0 = (1 + \varepsilon \sum Y_{\Sigma}^2(x))^{-1/2}, \quad (2)$$

where the mode shape of the cantilever mass sensor under tension is given by[24]

$$Y(x) = \sinh q_1 x - G_C \cosh q_1 x - q_1/q_2 \sin q_2 x + G_C \cos q_2 x \quad (3a)$$

and for suspended configuration the corresponding mode shape yields

$$Y(x) = \sinh q_1 x - G_B \cosh q_1 x - q_1/q_2 \sin q_2 x + G_B \cos q_2 x, \quad (3b)$$

where $x = X/L$, $G_C = (q_1^2 \sinh q_1 + \gamma_0^2 \sin q_2)/(q_1^2 \cosh q_1 + q_2^2 \cos q_2)$, $G_B = (\sinh q_1 - q_1/q_2 \sin q_2)/(\cosh q_1 - \cos q_2)$, $q_{1,2} = [\pm b^2/2 + (b^4/4 + \gamma_0^4)^{1/2}]^{1/2}$ and $b = (F_T L^2/EI)^{0.5}$ is the dimensionless tension parameter[20,25] and $EI$ is the bending stiffness. We must emphasis here that in Eq. (2) just the normalized mode shapes are used, i.e. $\int_0^1 Y^2(x)dx = 1$.

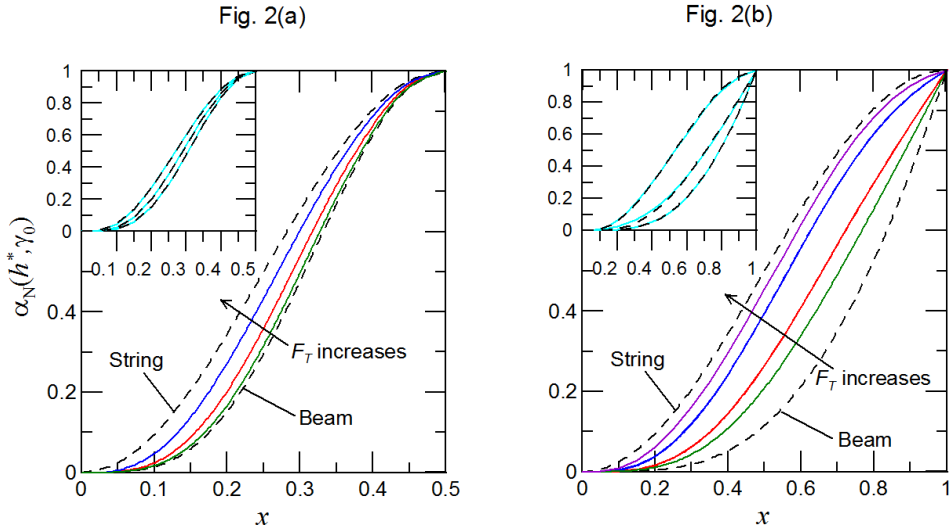

Fig. 2. The variation of $\alpha_N(h^*, \gamma_0)$ for different values of $F_T$, where $\alpha_N(h^*, \gamma_0) = \alpha(h^*, \gamma_0)/\alpha_{max}(h^*, \gamma_0)$ and $\alpha_{max}(h^*, \gamma_0)$ is the maximum value of $\alpha$, and for (a) suspended and (b) cantilever configuration, respectively. Insets present the comparison between the mode shapes (dash line), $Y_N(x) = [Y(x)/Y_{max}(x)]^2$ and position functions (solid line), $\alpha_N(h^*, \gamma_0) = \alpha(h^*, \gamma_0)/\alpha_{max}(h^*, \gamma_0)$, for different values of $F_T$. $Y_{max}(x)$ and $\alpha_{max}(h^*, \gamma_0)$ are maximum achievable values of mode shape and position function, respectively.

It is evident from structure of Eqs. (1) and (2) and also depicted in inset of Fig. 2 that the position function can be expressed through the mode shape and vice versa as follows

$$\alpha(h_i^*, \gamma_0) \approx (1/4)Y^2(x). \tag{4}$$

Notably, the following important conclusions can be drawn from Eq. (4): i) for cantilever configuration the expression for $\alpha(h_i^*, \gamma_0)$ is cumbersome (see Appendix), hence the energy approach could be preferred; and ii) method of multiple mass determination utilizing measurement of the cantilever several consecutive resonant frequencies[10] can be directly employed to quantitatively evaluate $N$ masses attached on the nanomechanical based mass sensor with intentionally applied axial tension. In this case, measurement of $P$ shifts of the fundamental resonant frequencies under different axial prestress forces (for beam $F_T = 0$) are used. Then, as in Dohn et. al.[10], Eqs. (1) and (2) can be rewritten in the following way

$$\mathbf{U}\boldsymbol{\rho}\,\overline{d} = \overline{R_\omega}, \tag{5}$$

where $\boldsymbol{\rho}\,\overline{d} = \begin{bmatrix} \varepsilon_1 \\ \varepsilon_2 \\ \vdots \\ \varepsilon_N \end{bmatrix}$, $\overline{d}$ is the *N*-elements unitary vector and for perturbation technique by Eq. (1)

$$\mathbf{U} = [\overline{\alpha_1}, \overline{\alpha_2}, ..., \overline{\alpha_N}], \overline{\alpha_N} = \begin{bmatrix} \sum_{j=1}^{m\Sigma} \alpha(h_j^*, \gamma_{01}) \\ \sum_{j=1}^{m\Sigma} \alpha(h_j^*, \gamma_{02}) \\ \vdots \\ \sum_{j=1}^{m\Sigma} \alpha(h_j^*, \gamma_{0P}) \end{bmatrix}, \overline{R_\omega} = \begin{bmatrix} \Delta f_1/(2f_{01}) \\ \Delta f_2/(2f_{02}) \\ \vdots \\ \Delta f_P/(2f_{0P}) \end{bmatrix},$$ whereas for the energy approach

by Eq. (2) $\mathbf{U} = [\overline{y_1}, \overline{y_2}, ..., \overline{y_N}], \overline{y_N} = \begin{bmatrix} \sum_{j=1}^{m\Sigma} Y_1^2(x_{h_j}) \\ \sum_{j=1}^{m\Sigma} Y_2^2(x_{h_j}) \\ \vdots \\ \sum_{j=1}^{m\Sigma} Y_P^2(x_{h_j}) \end{bmatrix}, \overline{R_\omega} = \begin{bmatrix} f_1/f_{01} \\ f_2/f_{02} \\ \vdots \\ f_P/f_{0P} \end{bmatrix}.$

Now we seek for positions $h_j^*$ and the relative mass changes $\varepsilon_i$ that satisfy the Eq. (5). Briefly, the expected attachment positions are obtained numerically by the method of least squares and afterwards for obtained $h_j^*$ the desired mass ratios $\varepsilon_i$ can be found; the reader is referred to Ref. 10 for detailed discussion on the numerical solution of Eq. (5).

Importantly, Refs. 21 and 22 and Fig. 2 indicate that to perform single and multiple masses determination by cantilever (suspended) based mass sensors requires tensile forces of $b > 3$ (5), i.e. within measured frequency shifts only one can be performed for lower values of *b*. Moreover, the accuracy of mass evaluation by axially loaded nanomechanical based mass sensors depends on the uncertainties in force measurement. Here, we use the perturbation technique to derive the relative error in mass determination $\Delta\varepsilon$ caused by the small uncertainties in force measurement represented through $\Delta\alpha$, $\Delta f$ and $\Delta f_0$, and, as a result, the one yields

$$\Delta\varepsilon/\varepsilon \ (= \Delta m/m) \approx \frac{f}{f_0}\left(\frac{\Delta f_0}{f_0 - f}\right) - \frac{\Delta f}{f_0 - f} - \frac{\Delta\alpha}{\alpha}, \qquad (6)$$

Resonant frequencies $f$ and $f_0$ are generally proportional to $b$ as $K_1 b^2 + K_2 b$, where $K_{1,2}$ are the coefficients depending on the resonator configuration and the applied axial force, e.g. for string regime $K_1 = 0$.[22] For cantilever (suspended) configuration of the mass sensor and $b \leq 3$ (5), variation of $\Delta\alpha/\alpha$ is negligible and the accuracy of mass measurement depends just and only on $\Delta f$ and $\Delta f_0$. In addition, typical force resolution of the micro-sized resonators is of sub-piconewton ($O(0.1\ \text{pN})$)[26] (in vacuum and for a low temperature the force accuracy attonewtons can be reached[27]) and for nano-sized resonators the detectible forces are of sub-femtonewton.[28]

## IV. DISCUSSION

In this section, the practicality of proposed method is demonstrated and its accuracy and sensitivity is evaluated. We consider suspended multi-walled carbon nanotube based mass sensor (MWCNT) of density 2.1 g/cm$^3$, elastic moduli 1.15 TPa, length of 11.4 μm with outer and innermost diameters of 15 nm and 3 nm, respectively; which is loaded by a molecule of $m = 122$ ag, i.e. $\varepsilon \approx 0.03$, with an attachment position at $h = 5.6$ μm. The uncertainty error in predetermined forces is 0.1 pN, i.e. three orders of magnitude higher than the usual force resolution of the nanomechanical resonators.[28] Then the fundamental resonant frequencies of the MWCNT under $F_T = 3.65$ and 16.7 nN are 5.403 MHz ± 0.06 kHz ($\gamma_0 \approx 7.021$) and 10.305 MHz ± 0.03 kHz ($\gamma_0 \approx 9.697$) and the attached mass causes the shift ($f_0\ f$) of 178.3 ± 0.1 and 320.3 ± 0.15 kHz, respectively. It gives the frequency shift ratio $[(f_{01} - f_1)/f_{01}]/[(f_{02} - f_2)/f_{02}]$ of 1.062 (i.e. error in this ratio is of one order lower, i.e. $O(10^{-4})$), where subscripts 1 and 2 stand for $F_T = 3.65$ and 16.7 nN, respectively. Accounting for dependence of $\alpha(h^*, 7.021)/\alpha(h^*, 9.697)$ obtained by solving equations given in Appendix, the value of 1.062 agrees with the possible attachment positions at $h^* = 0.49$ or $0.51$. Now for $\alpha(0.49/0.51, 7.021) = 0.57$ the desired mass ratio of $\varepsilon \approx 0.03$ is found yielding from a known mass of MWCNT the attached molecule mass of 122 ag. It is evident from this example that the commonly achievable uncertainties in force measurement[28] do not affect accuracy of the estimated mass(es).

For mass spectrometry in gaseous and aqueous solutions it is required that the dissipative effects due to surrounding fluid are small, i.e. $Q \gg 1$.[29] As a result, the resonant frequencies in Eqs. (1) and (2) can be replaced by those measured in gases and fluids, while the method of mass(es) determination is identical with the one in vacuum. Besides, fluid damping causes decrease of the mass sensitivity, whereas the axial tension its increase. Thus it is of practical importance to analyze an improvement of the mass sensitivity of nanomechanical based mass sensor in fluid caused by the applied tensile force. To begin we recall that in fluid and for $Q \gg 1$ the fundamental resonant frequencies are given by[29,30]

$$f_f = \frac{f_v}{\sqrt{1 + \frac{\pi \rho_l W_D^2}{4 \rho A} \Gamma_r(\omega_f)}}, \tag{7}$$

where $\rho$ and $A$ are the resonator density and cross sectional area, respectively; $W_D$ is dominant cross section scale (e.g. for MWCNT it is the outer diameter), $\rho_l$ is the fluid density and $\Gamma_r(\omega_f)$ is the real component of the hydrodynamic function, $\omega_f = 2\pi f_f$ and $f_v$ are the vacuum resonant frequencies.[21,22,25] Accounting for Eq. (7) the dimensionless mass sensitivity can be expressed as

$$S \approx (\gamma_0^2 - \gamma^2)/\varepsilon \left[1 - P\left(\frac{\rho_l}{\rho}\right)\Gamma_r(\omega_f)\right], \tag{8}$$

where $P = (\pi/8)(W/T)$ or $1/2$ for rectangular or circular resonator cross section, $T$ is the resonator thickness and $W$ is its width. Equation (8) and results given in Table 1 reveal that the required mass sensitivity, i.e. sensitivity needed to determine targeted molecule(s), can be achieved either by changing $F_T$ (for cantilever (suspended) resonator in fluid and $b \leq 3$ (5) the position functions and the mode shapes are identical with those known for a beam of without tension) or by optimizing the resonator dimension.

Table 1 The achievable mass sensitivities of cantilever resonator made of silicon of $L = 200$ μm, $W = 30$ μm and of thicknesses $T = 1, 2$ and $4$ μm in vacuum and air loaded by three different masses of the identical mass ratio $\varepsilon = 0.01$ at $h^* = 1$ as the function of applied axial tension.

|  | Vacuum | Air | | | |
|---|---|---|---|---|---|
|  | (all cases) | $T = 0.5$ μm | $T = 1$ μm | $T = 2$ μm | $T = 4$ μm |
| $b = 0$ | 6.84 | 6.68 | 6.76 | 6.80 | 6.82 |
| $b = 2$ | 10.54 | 10.30 | 10.42 | 10.48 | 10.51 |
| $b = 5$ | 15.96 | 15.61 | 15.78 | 15.87 | 15.91 |
| $b = 8$ | 20.01 | 19.59 | 19.80 | 19.91 | 19.96 |

## IV. CONCLUSIONS

We proposed a novel approach capable of the quantitative mass determination from only measured fundamental resonant frequencies of the micro-/nanomechanical based mass sensors. This method benefits from the mode shape and frequency shift changes caused by an intentionally axial tensile force. Our finding can find an application in a real-time mass spectrometry in gaseous and aqueous solutions, where usually just the fundamental resonant frequency can be correctly identified.[13] Moreover, since method requires measurement of only the fundamental modes, the impact of noise and damping on accuracy of the results is minimized.

In addition, it has been found that the commonly achievable uncertainties in force measurement have a negligibly small impact on the extracted mass values.

The author gratefully acknowledge the support provided to this research by Grant Agency of Czech Republic, GAČR 15-13174J

**APPENDIX**

Here we present the position function $\alpha_i(h_i^*, \gamma_0)$ of the $i$-th attached particle obtained by means of perturbation technique, which for a suspended configuration reads

$$\alpha_i(h_i^*, \gamma_0) = -H_{Di}(q_0, h_i^*)/[4A_1 H_{DT}(\gamma_0)], \tag{A1}$$

and for cantilever one it is described by

$$\alpha_i(h_i^*, \gamma_0) = -H_{Ci}(q_0, h_i^*)/[4A_1^4 H_{DC}(\gamma_0)], \tag{A2}$$

where

$H_{Bi}(q, h_i^*) = A_0 A_1 \sinh q_1 h_i^* \sinh (q_1(1 - h_i^*)) \sin q_2 - \sinh q_1 \sin q_2 h_i^* \sin (q_2(1 - h_i^*)) + A_0[F(q) - F(qh_i^*) - F(q(1 - h_i^*))] + A_1[G(q) - G(qh_i^*) - G(q(1 - h_i^*))] + A_1^2[F(qh_i^*) + F(q(1 - h_i^*)) - \sinh q_1 \cos q_2 h_i^* \cos (q_2(1 - h_i^*))] + A_0/A_1[G(qh_i^*) + G(q(1 - h_i^*)) - \cosh q_1 h_i^* \cosh (q_1(1 - h_i^*))\sin q_2]$, (A3)

$H_{Ci}(q, h_i^*) = A_1^4 F(qh_i^*) + A_1^3[\cosh q_1 h_i^* \sin (q_2(1-h_i^*)) \cos q_2 + \sinh q_1 h_i^* \sinh (q_1(1-h_i^*)) \sin q_2] + A_1^2[F(q) - F(q(1-h_i^*)) - \cosh q_1 h_i^* \cos (q_2(1-h_i^*)) \sin q_2] + A_1 G(q(1-h_i^*)) + A_1^5 A_2 \sin q_2[\cos q_2 + \sin q_2 h_i^* \sin (q_2(1-h_i^*))] + A_1^4 A_2[F(q(1-h_i^*)) - (\sinh q_1 + \sin q_2) \cos (q_2(1-h_i^*)) \cos q_2 h_i^*] + A_1^3 A_2[G(q) + \cosh (q_1(1-h_i^*)) \sin q_2 h_i^* \cos q_2] - A_1^2 A_2 [F(qh_i^*) + \sinh q_1 \sin q_2 h_i^* \sin (q_2(1-h_i^*)) + \cosh (q_1(1-h_i^*)) \cos q_2 h_i^* \sin q_2] + A_1 A_2 G(qh_i^*) - \cosh q_1 h_i^* \cosh (q_1(1-h_i^*)) \sin q_2 + (b/q_2)^2 \{A_1^2[F(qh_i^*) - \sinh (q_1 h_i^*) \cosh (q_1(1-h_i^*)) \cos q_2] + A_1[G(q) + G(qh_i^*) - G(q(1-h_i^*))] + A_1^3 A_2[G(q(1-h_i^*)) - \cosh q_1 \cos (q_2 h_i^*) \sin (q_2(1-h_i^*))] + A_1^2 A_2[G(q(1-h_i^*)) + G(qh_i^*) - G(q)] +$

$A_1A_2[\cosh q_1 \sin(q_2 h_i^*)\cos(q_2(1-h_i^*)) - G(qh_i^*)] + \cosh(q_1 h_i^*)\sinh(q_1(1-h_i^*))\cos q_2 - F(q(1-h_i^*))\}$, (A4)

$H_{DT}(\gamma_0) = [(2q_1^2 + b^2)/(2q_1q_2)]G(z_0) + b^2(3q_1^2 + q_2^2)/A_3 \sinh q_1 \sin q_2 - [(2q_2^2 - b^2)/(2q_2^2)]F(z_0) - [(8q_1^2 q_2^2 + q_1^4 + 3q_2^4)/A_3](\cosh q_1 \cos q_2 - 1)]$, (A5)

$H_{DC}(\gamma_0) = 2b^4/(A_4 A_5)(1 - \cosh q_1 \cos q_2) - 2(2q_1/A_4 + A_4/A_5)(\cosh q_1 \cos q_2 + 1) - [1 + b^2/(2q_2^2) + b^4/(2q_1 A_5)]\sinh q_1 \cos q_2 + [1/A_1 - b^2/(2q_1 q_2) + b^4/(2q_2 A_5)]\cosh q_1 \sin q_2 - [A_4/(q_1 q_2 A_5) + 4/(q_2 A_4)]\sinh q_1 \sin q_2$, (A6)

where $A_0 = (b/q_1)^2 - 1$, $A_1 = q_2/q_1$, $A_2 = [(b/q_1)^2 + A_1^2]/A_0$, $B_0 = (b/q_2)^2 - 1$, $B_1 = q_1/q_2$, $F(z) = \sinh z_1 \cos z_2$ and $G(z) = \cosh z_1 \sin z_2$, $A_3 = q_1^2 q_2(q_1^2 + q_2^2)$, $A_3 = q_1^2 q_2 A_4$, $A_4 = q_1^2 + q_2^2$ and $A_5 = q_1 q_2^2$.